\documentclass[preprint,5p]{elsarticle}
\usepackage{upgreek}
\usepackage{ragged2e}
\usepackage{amsmath}
\usepackage{lineno,hyperref}
\usepackage{amssymb}
\usepackage{graphicx,natbib}
\usepackage{ulem} 
\usepackage{tabularx}

\modulolinenumbers[250]

\begin{document}

\begin{frontmatter}

\title{Role of Neutron Transfer in Sub-Barrier Fusion}

\author[mymainaddress]{Rudra N. Sahoo}
\address[mymainaddress]{Department of Physics, Indian Institute of Technology Ropar, Rupnagar - 140001, Punjab, India}
\author[mymainaddress]{Malika Kaushik}
\author[mymainaddress]{Arshiya Sood}
\author[mymainaddress]{Pawan Kumar}
\author[mymainaddress]{Swati Thakur}
\author[mymainaddress]{Arzoo Sharma}
\author[mymainaddress]{Pushpendra P. Singh}
\author[mysecondaryaddress1]{Md. Moin Shaikh}
\address[mysecondaryaddress1]{Variable Energy Cyclotron Centre, 1/AF, Bidhannagar, Kolkata - 700064, India}
\author[mysecondaryaddress2]{Rohan Biswas}
\address[mysecondaryaddress2]{Nuclear Physics Group, Inter-University Accelerator Centre, New Delhi - 110 067, India}
\author[mysecondaryaddress3]{Abhishek Yadav}
\address[mysecondaryaddress3]{Department of Physics, Jamia Milia Islamia, New Delhi - 110025, India}
\author[mysecondaryaddress4]{Manoj K. Sharma}
\address[mysecondaryaddress4]{Department of Physics, Shri Varshney College, Aligarh, UP - 202001, India}
\author[mysecondaryaddress2]{J. Gehlot}
\author[mysecondaryaddress2]{S. Nath}
\author[mysecondaryaddress2]{N. Madhvan}

\begin{abstract}
Fusion excitation function of $^{35}$Cl + $^{130}$Te system is measured in the energy range around the Coulomb barrier and analyzed in the framework of the coupled-channels approach. The role of projectile deformation, nuclear structure, and the couplings of inelastic excitations and positive Q$-$value neutron transfer channels in sub-barrier fusion are investigated through the comparison of reduced fusion excitation functions of $^{35,37}$Cl +$^{130}$Te systems. The reduced fusion excitation function of $^{35}$Cl + $^{130}$Te system shows substantial enhancement over $^{37}$Cl + $^{130}$Te system in sub-barrier energy region which is attributed to the presence of positive Q-value neutron transfer channels in $^{35}$Cl + $^{130}$Te system. Findings of this work strongly suggest the importance of +2$n$ - transfer coupling in sub-barrier fusion apart from the simple inclusion of inelastic excitations of interacting partners, and are in stark contrast with the results presented by Kohley \textit{et al.}, [Phys. Rev. Lett. 107, 202701 (2011)].
\end{abstract}

\end{frontmatter}
\linenumbers

\section{Introduction}\label{1}
Sub-barrier fusion in heavy-ion induced reactions offers possibilities to explore static and dynamic properties of nuclei and to investigate the advancement of tunneling phenomena in terms of couplings of inelastic excitations and transfer channels \cite{bbback,am,lemasson,dasgupta2007,jiang2014,gm,gc}. Further, the understanding of sub-barrier fusion leads to the comprehensive knowledge of suitable conditions for the synthesis and exploration of superheavy elements, source of energy in astrophysical objects, lower breakup threshold of weakly bound nuclei, and fusion reaction dynamics at extreme low energies \cite{viz1,love,lfc1,mss,clj1,mdms,clj12}. Generally, the fusion of two heavy nuclei occurs if the entrance channel can overcome the effective barrier formed due to the cumulative effect of repulsive Coulomb and attractive nuclear potential. However, fusion at sub-barrier energies has been experimentally ascertained in different reports \cite{clj1,rgs1,wong,stefanini1}, and showed substantial enhancement over the standard one-dimensional barrier penetration model (1-D BPM) \cite{liang2003}. In numerous existing studies, the sub-barrier fusion has been attributed to the quantum tunnelling \cite{lcua,wong}, static deformations, dynamic deformations leading to the coupling of in-elastic excitations \cite{rgs1,stefanini1,kha1}, and due the onset of transfer channels \cite{dasgupta2007,vvs1}. 

Beckerman \textit{et al.} first observed the effect of positive Q$-$value neutron transfer channels in sub-barrier fusion $^{58}$Ni +$^{58}$Ni, $^{64}$Ni +$^{64}$Ni and $^{58}$Ni + $^{64}$Ni systems \cite{mb1}. The excitation functions of former two systems at sub-barrier energies have been found to be identical within the experimental uncertainties. For $^{58}$Ni +$^{64}$Ni system, the excitation function showed large enhancement as compared to the former two systems. Albeit all intrinsic properties of these systems are same, the presence of positive Q$-$value +2$n$ transfer channel in $^{58}$Ni +$^{64}$Ni system is an exception which has been correlated with sub-barrier fusion enhancement. Subsequently, the presence of positive Q-value neutron transfer channels in $^{16,18}$O + $^{60,58}$Ni, $^{28}$Si +$^{90,94}$Zr \cite{sk1}, $^{32}$S +$^{58,64}$Ni, $^{90,94,96}$Zr \cite{phs,hmj,hqz}, and $^{40,48}$Ca + $^{124,132}$Sn, $^{90,96}$Zr \cite{jjk1,ht1,AMS13} systems has been correlated with the sub-barrier fusion enhancement. However, the role positive Q$-$value neutron transfer channels in sub-barrier fusion is found to be negligible for $^{130}$Te +$^{58,64}$Ni \cite{zk1}, $^{60,64}$Ni + $^{100}$Mo \cite{ams1}, $^{132}$Sn + $^{58}$Ni \cite{zk1} and $^{64}$Ni + $^{118}$Sn \cite{ktl1}, $^{16,18}$O +$^{76,74}$Ge, $^{92}$Mo,$^{118}$Sn \cite{cjl1,mbwg,pj1}, and $^{40}$Ar + $^{112,122}$Sn \cite{wr} systems. Further, for $^{40}$Ar +$^{144,148,154}$Sm \cite{wr}, $^{46,50}$Ti +$^{124}$Sn \cite{jfl12}, $^{32,36}$S +$^{110}$Pd \cite{AMS12}, and $^{40,48}$Ca +$^{48}$Ca \cite{mt1} systems, the sub-barrier fusion enhancement has been interpreted by considering a combined effect of both positive Q$-$value neutron transfer channels and deformations. The ambiguity in observations suggests that the coupling of neutron transfer channels may not be sufficient but is of high importance to describe the sub-barrier fusion enhancement \cite{var,glz1}.

Several dynamical models have been proposed to interpret the sub-barrier fusion dynamics. In coupled-channels approach, Hagino \textit{et al.} \cite{kha1} included the coupling of positive Q$-$value +2$n$ transfer channel to explain the enhancement of sub-barrier fusion. Zagrebaev \textit{et al.} \cite{viz1,viz2} formulated a model by incorporating neutron transfer channels and used semi-classical approximation for transfer probability. Sargsyan \textit{et al.} \cite{vvs3} applied quantum diffusion approach to analyse fusion excitation functions of $^{11}$Be +$^{209}$Bi and $^{15}$C +$^{232}$Th reactions. A phenomenological model has been given by Rowley \textit{et al.} \cite{nrwl} by imposing transfer channels on coupled-channels calculations. Stelson \textit{et al.} \cite{phs} emphasized that the fusion occurs due to neutron flow during the process of interaction. Esbensen \textit{et al.}, numerically calculated sub-barrier fusion cross-sections of a few systems by including dynamical deformations in the coupled-channels calculations \cite{he1}. It has been found that the coupled-channels calculations do not reproduce the cross-sections of some systems, particularly because of probable transfer channels. The coupling of neutron transfer has been included in the calculations to interpret sub-barrier fusion \cite{he2,he3}. Pollorolo and Winther proposed a semi-classical approximation to explain the experimental fusion cross-sections \cite{gpaw}, which suggests that the transfer probability is quite small as compared to large projectile energy dissipation. Apart from these models, the universal fusion function \cite{vvs1} and the Wolki energy scaling law \cite{rw} have been proposed to interpret sub-barrier fusion data. Despite the existing studies, the role of different couplings in sub-barrier fusion enhancement is not yet fully understood. 

In this Letter, the role of entrance channel parameters in sub-barrier fusion is explored. The fusion excitation function for $^{35}$Cl + $^{130}$Te system is measured around the Coulomb barrier energies and analysed in the framework of coupled-channels calculations using theoretical model code CCFULL\cite{kha1}. Findings of the present work have been compared with that reported in ref.\cite{zk1} for $^{58,64}$Ni + $^{130}$Te systems and in ref.\cite{rns} for $^{37}$Cl + $^{130}$Te system to probe the role of transfer channels in sub-barrier fusion enhancement. The comparison of $^{35}$Cl + $^{130}$Te and $^{37}$Cl + $^{130}$Te systems is particularly interesting because $^{35}$Cl + $^{130}$Te system has six positive Q-value neutron transfer channels with respect to none in $^{37}$Cl + $^{130}$Te systems. The Q-value data of neutron transfer channels in $^{35,37}$Cl +$^{130}$Te systems are given in Table - \ref{tab:table2}.

\section{Experimental set-up and procedures}\label{2}
The experiments were performed at the Inter-University Accelerator Centre, New Delhi by employing a recoil mass separator, Heavy Ion Reaction Analyser (HIRA) \cite{aks}. The set-up and methodology are detailed in ref.\cite{rns}. However, a brief account of experimental conditions which are unique to this work is presented here. The $^{35}$Cl beams of energies E$_{\textrm{c.m.}}$ = 94.04 - 121.65 MeV, \textit{i.e.,} from 10$\%$ below to 15 $\%$ above the Bass barrier V$_{B}$ = 105.14 MeV, were bombarded onto a $^{130}$Te target of thickness $\approx$ 200 $\mu$g/cm$^{2}$ \cite{tarudra}. A $^{130}$Te target fabricated on a carbon backing of 20 $\mu$g/cm$^{2}$ was mounted in beam facing the carbon foil configuration inside the target chamber of HIRA maintained at a 10$^{-6}$ mbar vacuum. Two silicon surface barrier detectors each having 1 mm diameter aperture were mounted at 9.8 cm distance from the beam interaction point on the target foil subtending an angle of $\pm$ 15$^{0}$ on either side of the beam direction to monitor the beam. The evaporation residues (ERs) have been detected at the focal plane of HIRA through a multi-wire proportional counter of an active area of 150$\times$50 mm$^{2}$. The acceptance of HIRA was kept at 5 msr, 2.2$^\circ$ polar angle, and the transmission efficiency has been determined using the semi-microscopic Monte Carlo code TERS \cite{snath}.

\begin{figure}
\centering
\includegraphics[width=9.2cm,height=8.6cm, trim={4.5cm 0cm 1cm, 2 cm}]{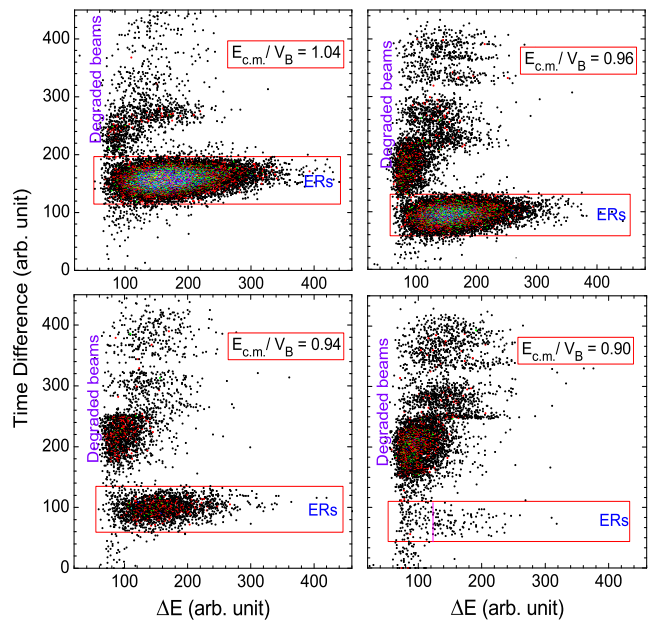}
\caption {\label{fig:spectra}(Color online) $\Delta$E -- Time spectra obtained for $^{35}$Cl + $^{130}$Te system at different incident energies, E$_\textrm{c.m.}$ / V$_\textrm{B}$ = 0.90, 0.94, 0.96 and 1.04.}
 \end{figure} 
   
The pulsed beams with a repetition rate of 2 $\mu$sec were used to achieve a clear separation between ERs and degraded beam. The time interval between two successive pulses was estimated to be greater than the flight time of ERs ($\approx$ 1.5 $\mu$sec at the energy around the barrier) during the passage through the dispersive elements of HIRA. The ERs have been identified by making an electronic gate between ToF and corresponding energy loss ($\Delta$E) through the multi-wire proposal counter. As a representative case, a few $\Delta$E -- Time spectra are shown in Fig.\ref{fig:spectra} where the ERs are well separated from the beam-like particles. However, at lowest measured energy, i.e., E$_\textrm{c.m.}$ / V$_\textrm{B}$ = 0.90, the ERs are not separated from degraded beam-like particles, particularly below channel number 125 on the x-axis (marked by a vertical pink line within the specified gate). In order to estimate the correct number of ERs in the region below channel number 125 on the x-axis, the number of counts in this region has been normalized between these two areas by comparing with the ratio of counts at higher energies (where the ERs are separated from degraded beam-like particles). The fusion cross-sections have been calculated using a standard formulation given in ref.\cite{rns}, and are presented in Table - \ref{tab:table1}. The cross-sections presented in Table - \ref{tab:table1} are the ERs cross-sections as the fission contribution in $^{35}$Cl +$^{130}$Te system is predicted to be negligible by theoretical model code PACE4 \cite{ag}. The errors presented in this table are absolute errors consisting of the statistical error and error in the transmission efficiency of HIRA.
 
\begin{figure} [h]
  	\centering
  	\includegraphics [trim={2cm 0cm 0cm 0cm},clip,width=15cm,height=12cm] {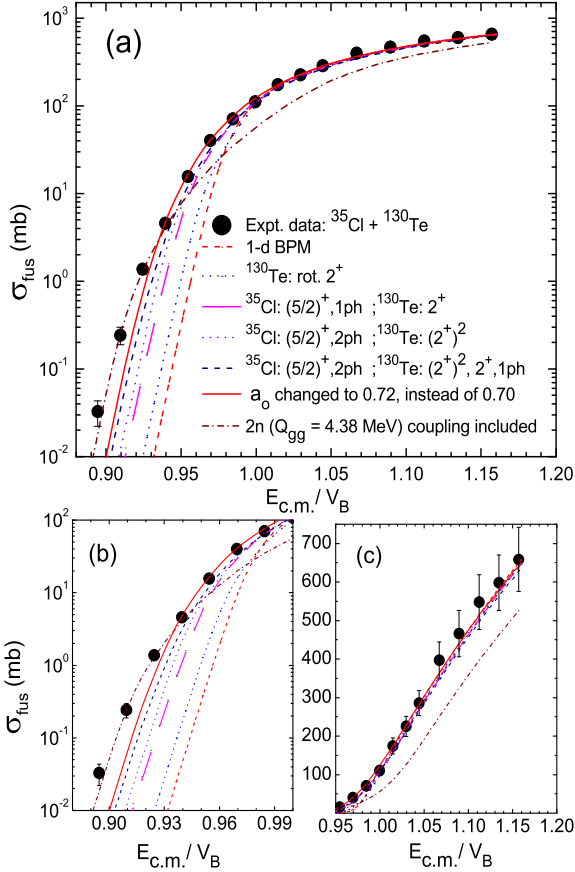}
  	\caption {\label{fig:ccfull}(Color online) $(a)$ The fusion excitation function of $^{35}$Cl + $^{130}$Te system with the outcome of coupled-channels calculations performed by including the couplings of in-elastic excitations of interacting partners, \textit{i.e.}, (5/2)$^{+}$ state of projectile with $\beta_{2}$=0.24, and (2$^{+}$)$^2$ and 2$^{+}$ (1 phonon) state of target with $\beta_{2}$=0.11 \cite{moller,nndc,rns}, and +2n transfer channel. $(b)$ Role of +2n transfer channel in sub-barrier energy region is highlighted. $(c)$ Fusion cross-sections at above barrier energies on linear scale for better visualization of the comparison between experimental data and coupled-channels calculations. The lines and symbol are self-explanatory, and the legend shown in $(a)$ also applies to $(b)$ and $(c)$.}
\end{figure}

\begin{table}
\caption{\label{tab:table2}The g.s.$\rightarrow$ g.s. Q -- value of neutron transfer channels in $^{35,37}$Cl + $^{130}$Te systems.}
\begin{tabular}{llllll} \hline
 +1n & +2n & +3n & +4n & +5n & +6n\\ \hline
$^{35}$Cl+$^{130}$Te & & & & & \\
+0.61 & +4.38 & +1.71 & +3.49 & +0.21 & +1.46 \\
$^{37}$Cl+$^{130}$Te & & & & & \\
-2.31 & -0.32 & -3.27 & -1.74 & -5.18 & -4.26  \\ \hline
\end{tabular}
\end{table}

\begin{table}[h!]
\caption{\label{tab:table1}The fusion cross-sections of evaporation residues (ERs) measured in $^{35}$Cl + $^{130}$Te system at different energies.} 
\begin{tabular}{llll} \hline
 E$_{\textrm{c.m.}}$(MeV)  & $\sigma_{\textrm{fus}}$(mb)  & E$_{\textrm{c.m.}}$(MeV)  & $\sigma_{\textrm{fus}}$(mb) \\ \hline
  94.0  & 0.034$\pm$0.010  & 106.7   & 174$\pm$21\\
  95.6  & 0.224$\pm$0.045  & 108.2   & 225$\pm$26 \\
  97.2  & 1.37$\pm$0.17    & 109.8   & 286$\pm$33 \\
  98.8  & 4.5$\pm$0.6     & 112.2   & 397$\pm$46\\
 100.3  & 15.6$\pm$1.8     & 114.5   & 466$\pm$60 \\
 101.9  & 40.2$\pm$4.5     & 116.9   & 548$\pm$70  \\                     
 103.5  & 71$\pm$8         & 119.3   & 598$\pm$72 \\
 105.0  & 111$\pm$13       & 121.6   & 659$\pm$83 \\ \hline    
\end{tabular}
\end{table}

\section{Results and coupled-channels analysis}\label{2}
Fig.\ref{fig:ccfull} shows fusion excitation function of $^{35}$Cl + $^{130}$Te system along with the outcome of coupled-channels calculations performed using theoretical model code CCFULL with standard Woods Saxon parametrization of nuclear potential \cite{kha1}, \textit{i.e.}, V$_{0}$ = 79.40 MeV, r$_{0}$ = 1.2 fm, and a$_{0}$ = 0.70 fm \cite{ams,hagino}. It may be pointed out that the CCFULL code does not consider the odd-even value of protons and neutrons. It only takes into account the excitation energy of low lying excitations, multi-polarity of the states, the value of deformation parameters, and the number of phonons to include more inelastic excitations states. As shown in Fig.\ref{fig:ccfull}, the calculations with 1-d BPM provide an excellent description of fusion cross-section above the barrier but underpredict at sub-barrier energies. In order to interpret sub-barrier fusion enhancement, the inelastic excitations of projectile ($^{35}$Cl) and target ($^{130}$Te) nuclei with modified diffuseness parameter (a$_{0}$=0.72 fm) are included in coupled-channels calculations. It can be noticed from Fig.\ref{fig:ccfull}, the revised coupled-channels calculations quantitatively well predict the fusion excitation function down to 6.65 $\%$ below the barrier energies but fails at lower energies. This discrepancy points towards the onset of some physical effect which is not implemented in the calculations. Since $^{35}$Cl + $^{130}$Te system has six positive Q-value neutron transfer channels, a pair of neutrons is included in the coupled-channels calculations, and the outcome of +2n transfer channel coupling is presented in Fig.\ref{fig:ccfull}. It may be pointed out that the inclusion of transfer coupling needs a transfer form factor (F$_\textrm{tra}$) derived from the transfer probability measured experimentally. In the present work, since the transfer cross-sections have not been measured, the coupled-channels calculations are performed with various transfer form factors. A transfer form factor 0.4 MeV reproduces very well the experimentally measured cross-sections at sub-barrier energies. This points towards the importance of positive Q$-$value neutron transfer channels in sub-barrier fusion enhancement. Further, the inclusion of +2n transfer channel coupling under-predicts experimentally measured fusion excitation function at above barrier energies, which is similar to what has been observed in Refs. \cite{khushboo,jie}.

\section{Role of neutron transfer channels}\label{2}
\begin{figure} [h]
   	\centering
   	\includegraphics[trim={4.2cm 1cm 0cm 1.5cm},clip,width=10cm,height=8.5cm] {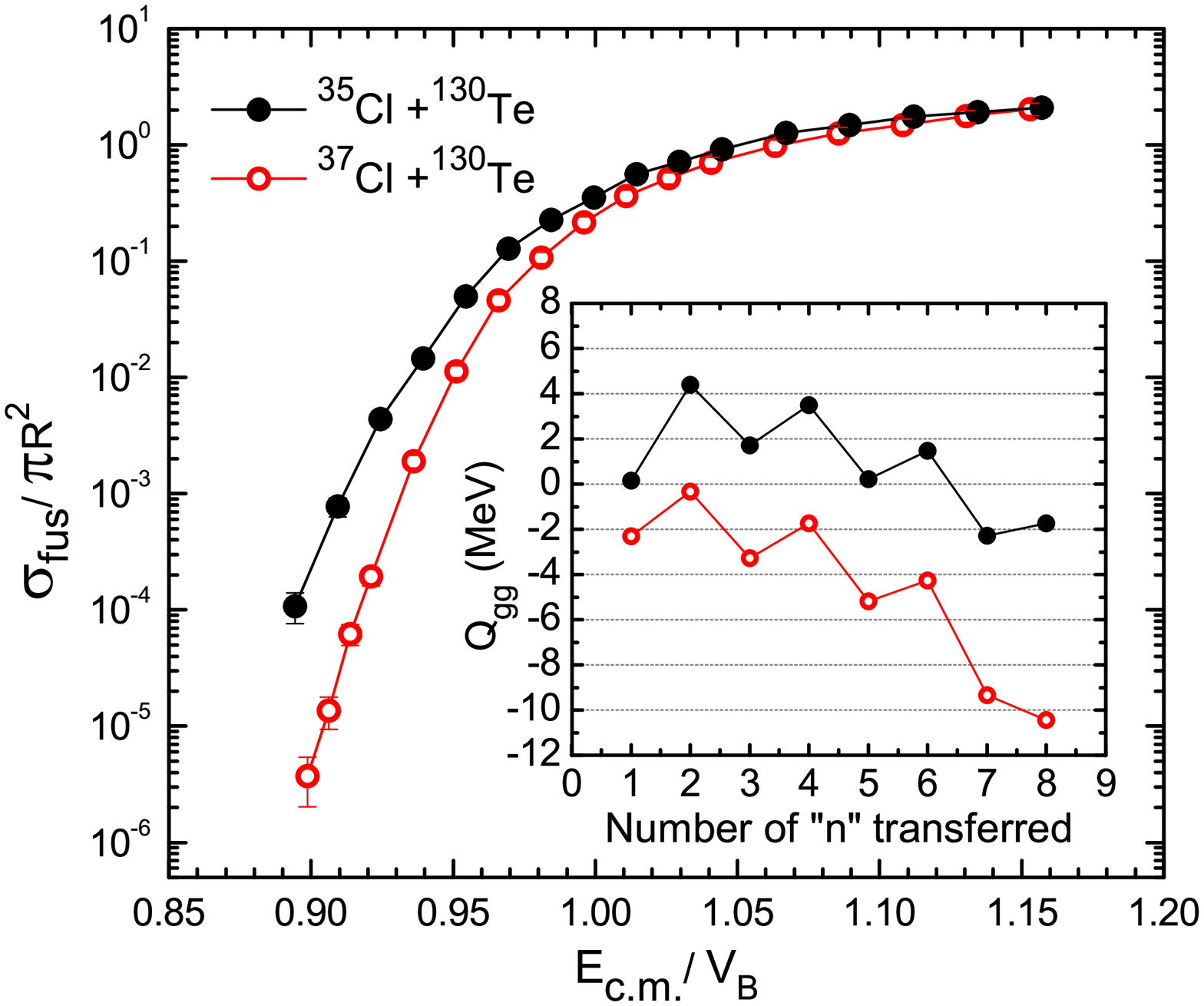}
   	\caption {\label{fig:comp}(Color online) The reduced fusion excitation functions of $^{35}$Cl + $^{130}$Te (this work) and $^{37}$Cl+$^{130}$Te \cite{rns} systems. Inset shows Q - values of neutrons transfer channels for both systems. Lines and curves are to guide the eyes.}
\end{figure}

In heavy-ion induced reactions, transfer of nucleons between interacting partners is highly regulated by optimum Q-value (Q$_{opt}$). Based on the value of Q$_{opt}$, the onset of all transfer channels is hindered including neutron stripping, but not the neutron pick-up and proton stripping \cite{lcorradi}. Among these two transfer process, the neutron pick-up is more probable compared to the proton stripping, because neutrons are insensitive to the coulomb barrier. For neutron transfer channels, the charge of projectile and target remains the same corresponds to zero optimum Q-value (Q$_{opt}$ = 0). Therefore, the ground state Q$-$value (Q$_{gg}$) is taken into consideration for the interpretation in the present work. To gain insights into the role of neutron transfer in sub-barrier fusion, reduced fusion excitation functions of $^{35}$Cl + $^{130}$Te (present work) and $^{37}$Cl + $^{130}$Te \cite{rns} systems have been compared in Fig.\ref{fig:comp}. It may be pointed out that the comparison of $^{35}$Cl and $^{37}$Cl data should be made on the same footings. Therefore, the center-of-mass beam energies (E$_{c.m.}$) and fusion cross-sections ($\sigma_{fus}$) are scaled with the respective Bass barriers (V$_{Bass}$) and geometrical cross-sections ($\pi \textrm{R}^{2}$) to remove effects of barrier position and nuclear radius of the two systems in comparison. The reduced fusion excitation function of $^{35}$Cl + $^{130}$Te system is found to be substantially higher than that of $^{37}$Cl + $^{130}$Te system in sub-barrier energy region, indicating a strong influence of projectile structure on sub-barrier fusion. From the geometric point of view, the fusion cross-sections in $^{37}$Cl + $^{130}$Te system at different energies are expected to be higher than $^{35}$Cl + $^{130}$Te system. Since the fusion cross-sections for both the systems are corrected by the normalization procedure mentioned above, the reduced excitation functions presented in Fig. \ref{fig:comp} should not display any significant difference within the experimental uncertainties. Fusion enhancement in case of $^{35}$Cl +$^{130}$Te system over $^{37}$Cl+$^{130}$Te system points toward the onset of neutron transfer channels provided that the fusion excitation function of $^{35}$Cl +$^{130}$Te system has been interpreted by including +2$n$ transfer channel as demonstrated in Fig.\ref{fig:ccfull}.

Further, the reduced excitation functions can be interpreted in terms of positive Q-value neutron transfer channels and spectroscopic properties of interacting partners. Note that the coupled-channel calculations satisfactorily reproduces the fusion excitation function of $^{37}$Cl+$^{130}$Te system by including couplings of low-lying excited states along with the modified barrier \cite{rns}. As given in Table - \ref{tab:table2}, $^{35}$Cl + $^{130}$Te system has six positive Q-value neutron transfer channels as compared to none in $^{37}$Cl+$^{130}$Te system. Sub-barrier fusion enhancement in case of $^{35}$Cl + $^{130}$Te system, refer to Fig.\ref{fig:comp}, is due to the presence of positive Q-value neutron transfer channels in this system. This suggests a strong correlation between sub-barrier fusion enhancement and the positive Q - value neutron transfer channels. Moreover, both $^{35}$Cl and $^{37}$Cl projectiles are deformed in different configurations with $\beta_{2}$ $\approx$ -0.24 and $\approx$ +0.11, respectively, and the excited states of the projectiles are of the nearly same order of magnitude. The oblate shape of $^{35}$Cl may reduce the effective barrier for fusion as compared to $^{37}$Cl, leading to the enhanced fusion cross-section for $^{35}$Cl+$^{130}$Te system in the sub-barrier region. 

In addition to this, the nuclear properties can influence low energy nuclear reactions as they are derived from the number of nucleons in the outermost valence shell of a nucleus. As shown in Fig.\ref{fig:comp}, the sub-barrier fusion enhancement in case of $^{35}$Cl +$^{130}$Te over $^{37}$Cl +$^{130}$Te system, maybe due to the isotopic effect of the projectiles. In $^{35}$Cl, the outermost shell, \textit{i.e.,} 1d$_{3/2}$ is half-filled, whereas, in $^{37}$Cl, it is fully filled. To attain stability, the $^{35}$Cl needs two extra neutrons to achieve filled outermost shell. During the interaction process between $^{35}$Cl and $^{130}$Te, neutrons may flow from $^{130}$Te to $^{35}$Cl due to configuration mixing between the interacting partners and may consequently fuse instead of exchanging two neutrons to achieve stability. In light of the qualitative picture, it may be inferred that the $^{35}$Cl show larger fusion probability as compared to $^{37}$Cl projectile with $^{130}$Te nucleus. This argument is in line of Stelson neutron flow model \cite{phs2}, but inconsistent with the observation that the heaviest isotopes of each element, which have the smallest neutron binding energies, give largest fusion cross-sections in the near barrier region \cite{phs1}.
  
\begin{figure} [h]
 	\centering
 	\includegraphics[trim={2cm 1.3cm 0cm 1.2cm},clip,width=12cm,height=9cm] {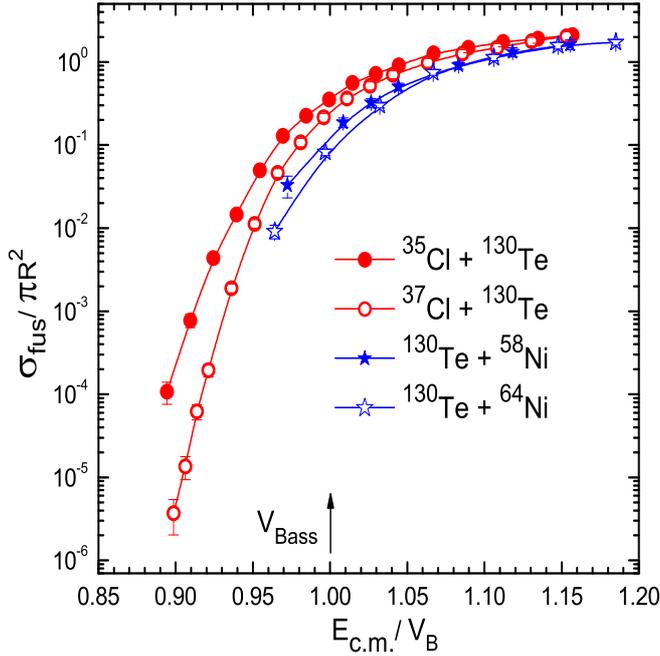}
 	\caption {\label{fig:NITE} (Color online) Comparison of reduced fusion excitation functions of $^{130}$Te + $^{58,64}$Ni \cite{zk1},  $^{35}$Cl + $^{130}$Te \cite{rns}, and $^{37}$Cl + $^{130}$Te (present work) systems.}
\end{figure}  

Recently, Kohley \textit{et al.,} \cite{zk1} reported that the presence of positive Q - value neutron transfer channels has very little, if any, influence on sub-barrier fusion. In ref. \cite{zk1}, the reduced fusion excitation functions of $^{58,64}$Ni + $^{130}$Te systems have been found to be equivalent even though $^{58}$Ni + $^{130}$Te system has 11 positive Q - value neutron transfer channels in comparison to only one in $^{64}$Ni +$^{130}$Te system. It has been concluded that the Te+Ni fusion measurements demonstrate lack of transfer effects. However, for $^{35,37}$Cl + $^{130}$Te systems, a strong correlation between positive Q - value neutron transfer channels and sub-barrier fusion has been observed. Fig. \ref{fig:NITE} shows reduced excitation functions of $^{58,64}$Ni + $^{130}$Te \cite{zk1} systems in comparison with $^{35,37}$Cl +$^{130}$Te systems. As can be noticed from this figure, the fusion cross-section measurements for $^{58,64}$Ni + $^{130}$Te systems were carried out down to the energy 3-4 $\%$ below the barrier. While in the present work, the fusion measurements are extended down to the energy 10 $\%$ below the barrier. It may be pointed out that the reduced excitation functions of $^{58,64}$Ni + $^{130}$Te \cite{zk1} and $^{35,37}$Cl + $^{130}$Te systems display almost identical behaviour down to the energy 3-4 $\%$ below the barrier. The splitting of excitation functions in case of $^{35,37}$Cl + $^{130}$Te systems is found to be more prominent at deep sub-barrier energies in a range from 4$\%$ to 10$\%$ down the barrier, indicating string influence of positive Q - value neutron transfer channels on sub-barrier enhancement. It would be interesting to investigate $^{58,64}$Ni + $^{130}$Te \cite{zk1} systems down to the deep sub-barrier energies for better insight into the role of positive Q - value neutron transfer channels in sub-barrier fusion.

\section{Summary and conclusions}\label{2}
The fusion excitation function for $^{35}$Cl + $^{130}$Te system has been measured from 10$\%$ below to 15$\%$ above the barrier and analyzed in the framework of coupled-channels approach. In conclusion, the influence of neutron transfer channels, especially +2n, and inelastic excitations couplings, have been observed. The reduced excitation functions of $^{35,37}$Cl + $^{130}$Te systems have been compared with the existing data of $^{58,64}$Ni + $^{130}$Te \cite{zk1} systems which have large variations in the number of positive Q-value nucleon transfer channels. It has been found that the reduced fusion excitation function of $^{35}$Cl + $^{130}$Te system displays substantial enhancement over $^{37}$Cl + $^{130}$Te system in sub-barrier energy region. The $^{35}$Cl + $^{130}$Te system has 6 positive Q-value nucleon transfer channels as compared to none in $^{35}$Cl + $^{130}$Te system. This strongly suggests that the positive Q-value nucleon transfer channels play an important role in sub-barrier fusion enhancement unlike the observations of Kohley \textit{et al.,} \cite{zk1} in which the fusion cross-sections were measured down to the energy 3-4 $\%$ below the barrier only.

Additionally, in the present work, the qualitative signature of the valence shell effect has been noticed for $^{35,37}$Cl + $^{130}$Te systems which are found to be in the line of the neutron flow model. Based on the results and interpretation presented in this Letter, it can be inferred that the sub-barrier fusion cross-sections are enhanced due to the superposition of tunneling, and the couplings of inelastic excitations and positive Q-value neutron transfer channels. It would be interesting to extend such measurements at deep sub-barrier energies for better insights into the role of deformation, nuclear structure, and transfer channels couplings for different systems. It may, however, be pointed out that a deeper understanding of sub-barrier fusion dynamics requires the identification of the contribution of the individual input parameter and the channel-by-channel cross-section measurement of transfer events.\\

{\bf{Acknowledgements}}\\

Authors acknowledge the accelerator staff of the Inter-University Accelerator Center, New Delhi for their efforts in delivering high quality $^{35}$Cl beams, the target lab personals for their help during the target fabrication, Z. Kohley for providing experimental data of their work \cite{zk1}, and R. G. Pillay, R. Chary and Vandana Nanal for their critical comments and suggestions during the preparation of this manuscript. One of the authors P.P.S. acknowledges a startup grant from the Indian Institute of Technology Ropar, and the Science and Engineering Research Board for Young Scientist Award No.YSS/2014/000250. \\

\end{document}